\documentclass[aps,prd,groupedaddress,showpacs,showkeys]{revtex4}
\usepackage{latexsym,epsfig,amsmath,amssymb}
\usepackage{amssymb}
\usepackage{amsmath}
\usepackage{amsfonts}
\usepackage{epsfig} 
\usepackage{verbatim} 

\newcommand{\seq}{\begin{subequations}}
\newcommand{\sen}{\end{subequations}}
\newcommand{\eq}{\begin{eqnarray}}
\newcommand{\en}{\end{eqnarray}}
\newcommand{\ra}{\rangle}

\def\lc{\Lambda_c(2940)^+}

\def\L2{\Lambda^2}

\begin{document}

\title{Strong two-body decays of the $\Lambda_c(2940)^+$ 
in a hadronic molecule picture}  

\noindent
\author{
        Yubing Dong$^{1,2}$, 
        Amand  Faessler$^3$,   
        Thomas Gutsche$^3$, 
        Valery E. Lyubovitskij$^3$\footnote{On leave of absence
        from Department of Physics, Tomsk State University,
        634050 Tomsk, Russia}
\vspace*{1.2\baselineskip}}
\affiliation{
$^1$ Institute of High Energy Physics, Beijing 100049, P. R. China\\ 
\vspace*{.4\baselineskip} \\
$^2$ Theoretical Physics Center for Science Facilities (TPCSF), CAS, 
Beijing 100049, P. R. China\\ 
\vspace*{.4\baselineskip} \\ 
$^3$ Institut f\"ur Theoretische Physik,  Universit\"at T\"ubingen,\\
Kepler Center for Astro and Particle Physics, \\ 
Auf der Morgenstelle 14, D--72076 T\"ubingen, Germany\\} 

\date{\today}

\begin{abstract} 

The $\lc$ baryon with possible quantum numbers $J^{\rm P} = \frac{1}{2}^+$ and
$\frac{1}{2}^-$ is studied as a
molecular state composed of a nucleon and $D^\ast$ meson. 
We give predictions for the strong two--body decay channels
$\lc \to pD^0$, $\Sigma_c^{++}\pi^-$ and $\Sigma_c^0 \pi^+$
where the sum of partial widths is consistent with current data
for the case of $J^{\rm P} = \frac{1}{2}^+$.
The case of $J^{\rm P} = \frac{1}{2}^-$ is shown to be ruled out.

\end{abstract}



\maketitle

\newpage

\section{Introduction}

Recently a new baryon resonance $\lc$ has been
discovered in the decay channel
$D^0 p$
by the {\it BABAR} Collaboration~\cite{Aubert:2006sp} and confirmed 
as a resonant structure in the final state
$\Sigma_c(2455)^{0,++} \pi^\pm \to \Lambda_c^+ \pi^+ \pi^-$ 
by Belle~\cite{Abe:2006rz}. Both collaborations report
on values for
the mass and width of the $\lc$ state which are consistent which
each other: 
$m_{\Lambda_c} = 2939.8 \pm 1.3 \pm 1.0$ MeV and 
$\Gamma_{\Lambda_c} = 17.5 \pm 5.2 \pm 5.9$ MeV 
({\it BABAR}~\cite{Aubert:2006sp}); 
$m_{\Lambda_c} = 2938.0 \pm 1.3^{+2.0}_{-4.0}$ MeV and 
$\Gamma_{\Lambda_c} = 13^{+8 \ + 27}_{-5 \ -7}$ MeV 
(Belle~\cite{Abe:2006rz}).

In Ref.~\cite{He:2006is} 
it was proposed that $\lc$ could be a $D^{\ast 0} p$ molecular state 
with spin--parity $J^{\rm P} = \frac{1}{2}^-$ or $\frac{3}{2}^-$, essentially
because its mass is just a few MeV below the $D^{\ast 0} p$ threshold value.
There it was also shown that the meson-exchange mechanism, such as $\pi$,
$\omega$ and $\rho$ exchange,
can lead to binding in such a $D^{\ast 0} p$ configuration.
In contrast, according to the predictions of the relativized quark
model the baryon state in the 2940 MeV mass 
region could have $J^{\rm P} = \frac{3}{2}^+$ or 
$\frac{5}{2}^-$~\cite{Capstick:1986bm}. A study of the decays
in the $^3P_0$ model~\cite{Chen:2007xf} excluded the possibility 
for $\lc$ to be the 
first radial excitation of the $\Lambda_c(2286)^+$ because the decay 
$\lc \to D^0 p$ vanishes in this case. The possibility however that $\lc$ 
is a $D$--wave charmed baryon with $J^{\rm P} = \frac{1}{2}^+$ or 
$\frac{3}{2}^+$ was shown to be favored.
In a relativistic heavy quark--light 
diquark model~\cite{Ebert:2007nw} it is suggested that the $\lc$ state  
could be the first radial $2S$ excitation of the $\Sigma_c(2455)^+$ 
with $J^{\rm P} = \frac{3}{2}^+$. 
It was also argued that the $\lc$ can result from
the first orbital excitation of the light diquark in the $\Lambda_c(2286)^+$.
An analysis of the strong decays of the $\lc$ in a chiral quark 
model~\cite{Zhong:2007gp} predicts that the $\lc$ is a $D$-wave state 
with principal quantum 
number $n=2$. In Ref.~\cite{Cheng:2006dk} the strong decays of charmed 
baryons have been studied in the framework of heavy hadron chiral 
perturbation theory (HHChPT). The conclusion of~\cite{Cheng:2006dk} 
on the nature of the $\lc$ was that an experimental determination
of the decay ratio 
$\Sigma_c^\ast\pi/\Sigma_c\pi$ will enable to discriminate the
$J^{\rm P}$ assignments. In Ref.~\cite{Gerasyuta:2007un} the $\lc$ state 
with $J^{\rm P} = \frac{1}{2}^-$ has been considered in the relativistic 
quark model. In Ref.~\cite{Roberts:2007ni} possible 
assignments ($J^{\rm P} = \frac{1}{2}^+$, $\frac{3}{2}^+$ and 
$\frac{5}{2}^-$) for the $\lc$ states have been analyzed in  
the quark model. The predicted masses for the $\lc$ state are distributed 
in the range from 2.887 GeV to 2.983 GeV. 
The coupled--channel approach~\cite{GarciaRecio:2008dp} does not 
generate any resonance in the energy region around 2940 MeV, which 
couples predominantly to the $ND^\ast$ pair having its threshold only a few 
MeV above the $\Lambda_c(2940)$ mass. Therefore, the conclusion of 
Ref.~\cite{GarciaRecio:2008dp} is that the $\Lambda_c(2940)$ is not 
a molecular $ND^\ast$ system.
Presently various possible structure interpretations exist
for the $\lc$ baryon, also depending on the particular model 
applied (see e.g. Refs.~\cite{Valcarce:2008dr,Chen:2009tm,Wang:2008vj}).  

In this paper we consider the $\lc$ as a molecular state composed of a nucleon
and a $D^\ast$ meson.
We also test the two currently possible assignments for the spin--parity
quantum numbers of $J^{\rm P} = \frac{1}{2}^+$ and $\frac{1}{2}^-$
in a molecular interpretation. Proceeding with this interpretation
the strong two--body decay modes
$p D^0$, $\Sigma_c^{++}\pi^-$ and $\Sigma_c^{0} \pi^+$ of the $\lc$
are evaluated. As one result we deduce that the assignment of
$J^P = \frac{1}{2}^+$ is favored in the molecular interpretation,
while the case $J^P = \frac{1}{2}^-$ gives an overestimate for
the total decay width. The technique for describing and treating
composite hadron systems we developed already in
Refs.~\cite{Faessler:2007gv,Dong:2008gb}. There we
present the formalism for the study of recently observed 
unusual meson states (like $D_{s0}^\ast(2317)$, $D_{s1}(2460)$, $X(3872)$, 
$Y(3940)$, $Y(4140$, $\cdots$) as hadronic molecules. 
The composite structure of these possible molecular states is
set up
by the compositeness condition $Z=0$~\cite{Weinberg:1962hj,%
Efimov:1993ei,Anikin:1995cf} 
(see also Refs.~\cite{Faessler:2007gv,Dong:2008gb}).  
This condition implies that the renormalization constant of the hadron 
wave function is set equal to zero or that the hadron exists as a bound 
state of its constituents. The compositeness condition was originally 
applied to the study of the deuteron as a bound state of proton and
neutron~\cite{Weinberg:1962hj}. Then it was extensively used
in low--energy hadron phenomenology as the master equation for the
treatment of mesons and baryons as bound states of light and heavy
constituent quarks (see e.g. Refs.~\cite{Efimov:1993ei,Anikin:1995cf}). 
By constructing a phenomenological Lagrangian including the 
couplings of the bound state to its constituents and the constituents 
to other final state particles we evaluated meson--loop 
diagrams which describe the different decay modes of the molecular states 
(see details in~\cite{Faessler:2007gv,Dong:2008gb}). 

In the present paper we proceed as follows. In Sec.~II
we first discuss the basic notions of our approach. We discuss the effective
Lagrangian for the treatment of the $\lc$ baryon 
as a superposition of the $p D^{\ast 0}$ and $n D^{\ast +}$ molecular 
components. Moreover, we consider the two--body hadronic 
decays $\lc \to p D^0, \Sigma_c^{++} \pi^-, \Sigma_c^0 \pi^+$ in this section. 
In Sec.~III we present our numerical results. 
Finally, in Sec.~IV we present a short summary of our results.

\section{Approach} 

In this section we discuss the formalism for the study of the
$\lc$ baryon. First we adopt the convention that the spin and parity 
quantum numbers of the $\lc$ are $J^{\rm P} = \frac{1}{2}^{+}$. 
But we also check the possibility of $J^{\rm P} = \frac{1}{2}^{-}$ (it 
will be shown that the last possibility leads to an overestimate of the
strong two--body decay widths and therefore must be excluded in 
our approach). 

Following Ref.~\cite{He:2006is} we consider this state as a superposition 
of the molecular $p D^{\ast 0}$ and $n D^{\ast +}$ components with 
the adjustable mixing parameter or mixing angle $\theta$: 
\eq\label{Xstate}
|\lc\ra =   \cos\theta \ | p D^{\ast 0} \ra \ 
        + \ \sin\theta \ | n D^{\ast +} \ra 
\, . 
\en 
The values $\sin\theta = 1/\sqrt{2}$, $\sin\theta = 0$ or  
$\sin\theta = 1$ correspond to the cases of ideal mixing, of 
a vanishing $n D^{\ast +}$ or $p D^{\ast 0}$ component, respectively. 
Our approach is based on an effective interaction Lagrangian describing 
the coupling of the $\lc$ to its constituents. We propose a setup for 
the $\lc$ in analogy to mesons consisting of a heavy quark and light antiquark,
i.e. the heavy $D^\ast$ meson defines the center of mass of the $\lc$, while 
the light nucleon surrounds the $D^\ast$. The distribution of the nucleon
relative to the $D^\ast$ meson we describe by the correlation function 
$\Phi(y^2)$ depending on the Jacobi coordinate $y$. 
The simplest form of such a Lagrangian reads 
\eq\label{Lagr_Lc}
{\cal L}_{\Lambda_c}(x) &=& g_{_{\Lambda_c}} \, 
\bar\Lambda_c^+(x) \, \Gamma^\mu \int d^4y \, \Phi(y^2) \, 
\Big( \cos\theta \, D^{\ast 0}_\mu(x) \, p(x+y) 
    + \sin\theta \, D^{\ast +}_\mu(x) \, n(x+y) \Big) \ + \ {\rm H.c.} 
\,, 
\en 
where $g_{_{\Lambda_c}}$ is the coupling constant of the $\lc$ to 
the constituents and $\Gamma^\mu$ is the corresponding Dirac matrix 
related to the spin--parity of the $\lc$. In particular we have 
$\Gamma^\mu = \gamma^\mu$ for $J^{\rm P} = \frac{1}{2}^+$ and 
$\Gamma^\mu = \gamma^\mu\gamma^5$ for $J^{\rm P} = \frac{1}{2}^-$. 
A basic requirement for the choice of an explicit form of the correlation 
function $\Phi(y^2)$ is that its Fourier transform vanishes sufficiently 
fast in the ultraviolet region of Euclidean space to render the Feynman 
diagrams ultraviolet finite. We adopt a Gaussian form for the correlation 
function. The Fourier transform of this vertex is given by
\eq 
\tilde\Phi(p_E^2/\Lambda^2) \doteq \exp( - p_E^2/\Lambda^2)\,,
\en 
where $p_{E}$ is the Euclidean Jacobi momentum. Here, 
$\Lambda \sim m_N \sim 1$ GeV is a size parameter characterizing 
the distribution of the nucleon in the $\lc$ baryon, which is of order of 
the nucleon mass or 1 GeV. The size parameter $\Lambda$ is 
a free parameter in our approach.

The coupling constant $g_{_{\Lambda_c}}$ is determined by the 
compositeness condition~\cite{Weinberg:1962hj,Efimov:1993ei,Anikin:1995cf,%
Faessler:2007gv}. It implies that the renormalization constant of 
the hadron wave function is set equal to zero with:
\eq\label{ZLc}
Z_{\Lambda_c} = 1 - \Sigma_{\Lambda_c}^\prime(m_{\Lambda_c}) = 0 \,.
\en
Here, $\Sigma_{\Lambda_c}^\prime(m_{\Lambda_c})$   
is the derivative of the $\lc$ mass operator shown in Fig.1. 
In order to evaluate the coupling $g_{_{\Lambda_c}}$ we use the
standard free propagators for the intermediate particles: 
\eq 
iS_N(x-y)=\left<0|TN(x)\bar N(y)|0\right>=\int\frac{d^4k}{(2\pi)^4i}\, 
e^{-ik(x-y)} S_N(k),\quad S_N(k)=
\frac{1}{m_N - \not\! k-i\epsilon}
\en 
for the nucleons and 
\eq
iS_{D^\ast}^{\mu\nu}(x-y)=
\left<0|TD^{\ast\,\mu}(x)D^{\ast\,\nu\,\dagger} (y)|0\right>=
\int\frac{d^4k}{(2\pi)^4i}\,e^{-ik(x-y)} S^{\mu\nu}_{D^\ast}(k)\,,
\quad S^{\mu\nu}_{D^\ast}(k)=
\frac{-g^{\mu\nu}+k^\mu k^\nu/m_{D^\ast}^2}{m_{D^\ast}^2-k^2-i\epsilon}
\en 
for the $D^\ast$ vector mesons. 

In order to study the strong two--body decays 
$\lc \to p D^0$ and $\lc \to \Sigma_c^{++,0}  \pi^{-,+}$ (see Fig.2) 
we need to know the coupling of the $\lc$ constituents to the final 
state particles. 
In particular, we need the Lagrangian describing the coupling of $ND^\ast$ 
pairs to $pD^0$ and $\Sigma_c\pi$. Such effective Lagrangians 
containing the coupling of two baryon fields with
$J^{\rm P} = \frac{1}{2}^+$, one vector and one pseudoscalar field have in 
general the form: 
\eq 
{\cal L}_{VPBB} \sim \bar B i\gamma^\mu \gamma^5 B V_\mu P  + {\rm H.c.} 
\,.  
\en 
The derivation of such a flavor SU(4) invariant Lagrangian is discussed 
in Appendix A. 
Here we just display the terms relevant for our calculations: 
\eq 
{\cal L}_{VPBB}(x) &=& - \frac{G}{F_D} 
\bar p(x) i \gamma^\mu \gamma^5 \biggl( 
  \frac{2}{5} p(x) D^{\ast 0}_\mu(x) 
+ \frac{1}{2} n(x) D^{\ast +}_\mu(x) \biggr) \bar D^0(x) \nonumber\\ 
&+&\frac{G}{F_\pi}
\bar \Sigma_c^{++}(x) i \gamma^\mu \gamma^5 \biggl( 
  \frac{9}{10} p(x) D^{\ast 0}_\mu(x) 
+ n(x) D^{\ast +}_\mu(x) \biggr) \pi^+(x) \\
&+&\frac{G}{F_\pi} \bar \Sigma_c^{0}(x) 
i \gamma^\mu \gamma^5 \biggl( p(x) D^{\ast 0}_\mu(x) 
+ \frac{9}{10} n(x) D^{\ast +}_\mu(x) \biggr) \pi^-(x) + \ {\rm H.c.} 
\nonumber
\en 
where $G = g_{\rho\pi\pi} g_A$ is the coupling constant; 
$g_{\rho\pi\pi} = 6$ is the $\rho\pi\pi$ coupling and
$g_A = 1.2695$ is the nucleon axial charge; 
$F_\pi = f_\pi/\sqrt{2} = 92.4$ MeV and $F_D = f_D/\sqrt{2} = 145.5$ MeV 
are the leptonic decay constants of $\pi$ and $D$ mesons, respectively. 

The strong two--body decay widths of the $\lc$ baryon are calculated 
according to the expressions: 
\eq\label{Gamma_PP}  
\Gamma(\Lambda_c[1/2^+] \to B + M) = 
\frac{g_{\Lambda_cBM}^2 }{16\pi m_{\Lambda_c}^3} \, 
\lambda^{1/2}(m_{\Lambda_c}^2,m_B^2,m_M^2) \, 
\Big( (m_{\Lambda_c} - m_B)^2 - m_M^2 \Big)
\en 
for the positive parity $\lc$ state and accordingly 
\eq\label{Gamma_NP} 
\Gamma(\Lambda_c[1/2^-] \to B + M) = 
\frac{f_{\Lambda_cBM}^2 }{16\pi m_{\Lambda_c}^3} \, 
\lambda^{1/2}(m_{\Lambda_c}^2,m_B^2,m_M^2) \, 
\Big( (m_{\Lambda_c} + m_B)^2 - m_M^2 \Big)
\en 
for the negative parity choice for $\lc$. The letters   
$B$ and $M$ denote the final baryon and pseudoscalar meson; 
$\lambda(x,y,z) = x^2 + y^2 + z^2 - 2xy - 2yz - 2xz$ is the K\"allen 
function; $m_{\Lambda_c}$,  $m_B$ and $m_M$ are the masses of the $\lc$, 
the final 
baryon $B$ and the meson $M$. In above expressions $g_{\Lambda_cBM}$ and 
$f_{\Lambda_cBM}$ are the effective coupling constants defining the 
interaction of the $\lc$ having quantum numbers
$J^{\rm P} = \frac{1}{2}^+$ or $\frac{1}{2}^-$ with the $(BM)$ pair: 
\eq\label{eff_couplings} 
{\cal L}_{\Lambda_cBM}^{1/2^+}(x) &=& g_{\Lambda_cBM} 
\bar\Lambda_c(x) i \gamma_5 B(x) M(x) + {\rm H.c.}\, \nonumber\\
{\cal L}_{\Lambda_cBM}^{1/2^-}(x) &=& f_{\Lambda_cBM} 
\bar\Lambda_c(x) B(x) M(x) + {\rm H.c.} \,.  
\en 

\section{Numerical results}

In Table 1 we present the numerical results for the partial two--body decay 
widths of the $\lc$ at $\Lambda = 1$ GeV -- the dimensional parameter
describing the distribution of the nucleon around the $D^\ast$ which is
located in the center-of-mass of the $\lc$. Note that the results are 
sensitive to the choice of the cutoff parameter $\Lambda$. An increase of  
$\Lambda$ leads to large values for the $\Lambda_c(2940)^+$ decay widths. 
We also vary $\cos\theta$ -- the mixing parameter 
of the $p D^{\ast 0}$ and $n D^{\ast +}$ components from 0 to 1.
For the case $J^{\rm P} = \frac{1}{2}^+$ the sum of the partial decay
widths is of the order of 1 MeV, at least consistent with current upper values
set by the observed total width. For the alternative case 
of $J^{\rm P} = \frac{1}{2}^-$ the dominant partial decay width is 
about 1 GeV in complete contradiction with the experimental constraints.
This dramatic increase in magnitude of the partial decay widths
for $J^{\rm P} = \frac{1}{2}^-$ is mainly explained by the large
phase space integral [see Eqs.~(\ref{Gamma_PP}) and (\ref{Gamma_NP})].   
We therefore conclude that in the context of a molecular interpretation
spin--parity $J^{\rm P} = \frac{1}{2}^+$ of the  $\lc$ state is clearly 
favored. The hadron molecule scenario with $J^{\rm P} = \frac{1}{2}^+$
results in partial decay widths for the modes $\Sigma_c^{++}\pi^-$
and $\Sigma_c^{0} \pi^+$, which are dominant and about equal.
The decay channel $p D^0$ is suppressed relative to $\Sigma_c^{++}\pi^-$
by a factor of about 4, details depending on the explicit value
of the mixing angle. Also for transparency we present results for 
the effective couplings $g_{\Lambda_cBM}$ and 
$f_{\Lambda_cBM}$ of Eq.~(\ref{eff_couplings}): 
\eq 
& &g_{\Lambda_c p D^0} = - 0.43 \pm 0.10\,, \hspace*{.2cm} 
g_{\Lambda_c \Sigma_c^{++}\pi^-} = 1.46 \pm 0.30\,, \hspace*{.2cm} 
g_{\Lambda_c \Sigma_c^{0}\pi^+} = 1.46 \pm 0.29\,, \nonumber\\
& &f_{\Lambda_c p D^0} = 1.26 \pm 0.34\,, \hspace*{.2cm} 
f_{\Lambda_c \Sigma_c^{++}\pi^-} = - 4.25 \pm 0.97\,, \hspace*{.2cm} 
f_{\Lambda_c \Sigma_c^{0}\pi^+} = - 4.42 \pm 0.78\,. 
\en  

\section{Conclusions} 

We pursue a hadronic molecule interpretation of the recently
observed $\lc$ baryon studying its consequences for the
the strong two--body decay modes and the $J^P$ quantum numbers.
In the present scenario the  $\lc$ baryon is supposed to be
described by a superposition of 
$|p D^{\ast 0}\ra$ and $|n D^{\ast +}\ra$ components with the
explicit admixture expressed by the mixing angle $\theta $.
The possible decay channels $pD^0$, $\Sigma_c^{++}\pi^-$ and
$\Sigma_c^0 \pi^+$ are fed by a $N D^{\ast}$ meson loop which
in turn arises from the hadronic constituents of the $\lc$.
The choice $J^{\rm P} = \frac{1}{2}^-$ is completely excluded
by the present calculation resulting in partial decay widths
of the order of 1 GeV.
For $J^{\rm P} = \frac{1}{2}^+$ we obtain the dominant
decay channels $\Sigma_c^{++}\pi^-$ and
$\Sigma_c^0 \pi^+$ relative to the $pD^0$ mode. The absolute rates
but less so the ratios of rates depend on the explicit molecule configuration
expressed by $\theta$. The sum of partial decay widths is consistent
with the upper value set by the observed total width.
An experimental determination of the partial decay
widths for the modes $pD$ and $\Sigma_c \pi$ could certainly help
in clarifying the structure issue involving the $\lc$ baryon.

\begin{acknowledgments}

This work was supported by the DFG under Contract No.
FA67/31-2 and No. GRK683.
This work is supported  by the National Sciences Foundations 
No. 10775148 and by the CAS grant No. KJCX3-SYW-N2 (YBD). 
This research is also part of the 
European Community-Research Infrastructure Integrating Activity
``Study of Strongly Interacting Matter'' (acronym HadronPhysics2,
Grant Agreement No. 227431) and President grant of Russia
``Scientific Schools''  No. 871.2008.2. 
The work is partially supported by Russian Science and Innovations 
Federal Agency under contract  No 02.740.11.0238. 
V.E.L. would like to thank the members of the Theory Group of
the Institute of High Energy 
Physics (Beijing) for their hospitality. 

\end{acknowledgments}

\appendix\section{Derivation of the phenomenological $VPBB$ 
interaction Lagrangian} 

First we consider the derivation of 
the phenomenological flavor SU(3) $VPBB$ interaction Lagrangian 
describing the coupling of vector ($V$) and pseudoscalar ($P$) mesons
to two baryons ($\bar B B$).  
It can be generated by starting with the ${\cal O}(p)$ term of 
chiral perturbation theory (ChPT) describing the coupling of baryon fields 
$(\bar B, B)$  with the chiral vielbein field $u_\mu$: 
 \eq 
{\cal L}_{PBB} = \frac{D}{2} {\rm tr} 
\Big(\bar B \gamma^\mu \gamma^5 \{u_\mu B \}\Big) 
 \, + \,   \frac{F}{2} {\rm tr} 
\Big(\bar B \gamma^\mu \gamma^5  [u_\mu B ]\Big) \, .
\en 
$D$ and $F$ are the baryon axial coupling constants 
(we restrict to the SU(3) symmetric limit, where $D = 3F/2 = 3g_A/5$ 
with $g_A = 1.2695$ being the nucleon axial charge); 
the symbols ${\rm tr}$, $\{ \ldots \}$ and $[ \ldots ]$ denote the trace 
over flavor matrices, anticommutator and commutator, respectively.
We use the standard notation for the basic blocks of the ChPT
Lagrangian~\cite{ChPT}, where $B$ is the octet of baryon fields, 
$U=u^{2}=\exp (iP\sqrt{2}/F_P)$
is the chiral field collecting pseudoscalar fields $P$
in the exponential parametrization with
$F_P$ being the octet leptonic decay constant, 
$u_{\mu }=iu^\dagger \nabla_\mu U u^\dagger$,  
$\nabla_{\mu }$ denotes the covariant derivative acting on the 
chiral field including external vector $(v_\mu)$ and axial $(a_\mu)$ 
sources: $\nabla_\mu U = \partial_\mu U - i (v_\mu + a_\mu) U 
+ i U (v_\mu - a_\mu)$.

The vector sources can 
be identified with the vector mesons $V_\mu$ if the latter are considered 
as gauge particles and introduced via the minimal substitution 
(for more details see e.g.~\cite{Bando:1987br}). The SU(3) baryon ($B$), 
pseudoscalar meson ($P$) and vector meson ($V$) matrices read as: 
\eq 
B = 
\left(
\begin{array}{ccc}
\Sigma^0/\sqrt{2} + \Lambda/\sqrt{6}\,\, & \,\, \Sigma^+ \,\, & \, p \\
\Sigma^- \,\, & \,\, -\Sigma^0/\sqrt{2}+\Lambda/\sqrt{6}\,\, & \,  n\\
\Xi^-\,\, & \,\, \Xi^0 \,\, & \, -2\Lambda/\sqrt{6}\\
\end{array}
\right),  
\en 
\eq 
P = 
\left(
\begin{array}{ccc}
\pi^0/\sqrt{2} + \eta/\sqrt{6}\,\, & \,\, \pi^+ \,\, & \, K^+ \\
\pi^- \,\, & \,\, -\pi^0/\sqrt{2}+\eta/\sqrt{6}\,\, & \, K^0\\
K^-\,\, & \,\, \bar K^0 \,\, & \, -2\eta/\sqrt{6}\\
\end{array}
\right),  
\en 
\eq 
V = 
\left(
\begin{array}{ccc}
\rho^0/\sqrt{2} + \omega/\sqrt{2}\,\, & \,\, \rho^+ \,\, & \, K^{\ast +} \\
\rho^- \,\, & \,\, -\rho^0/\sqrt{2}+\omega/\sqrt{2}\,\, & \, K^{\ast 0}\\
K^{\ast -}\,\, & \,\, \bar K^{\ast 0} \,\, & \, - \phi \\
\end{array}
\right). 
\en 
The required 
SU(3) $VPBB$ interaction Lagrangian reads: 
\eq 
{\cal L}_{VPBB} = g_V \frac{D+F}{F_P}  
{\rm tr} \Big(\bar B i \gamma^\mu \gamma^5 [V_\mu P] B\Big) 
+ g_V \frac{D-F}{F_P}  
{\rm tr} \Big(\bar B i \gamma^\mu \gamma^5 B [V_\mu P] \Big) \; , 
\en 
where $g_V = g_{\rho\pi\pi} = 6$ is the strong $\rho\pi\pi$ coupling
constant.
The extended SU(4) $VPBB$ interaction Lagrangian has the more complicated form 
\eq 
{\cal L}_{VPBB} &=& 
    ig_1 \bar B^{kmn} \gamma^\mu\gamma^5 [V_\mu, P]^l_k B_{lmn} 
 +  ig_2 \bar B^{kmn} \gamma^\mu\gamma^5 [V_\mu, P]^l_k B_{lnm}\nonumber\\ 
&+& ig_3 \bar B^{kmn} \gamma^\mu\gamma^5 ((V_\mu)^l_k P^s_m - 
                      P^l_k (V_\mu)^s_m) B_{lns} \\
&-& ig_3 \bar B^{knm} \gamma^\mu\gamma^5 ((V_\mu)^l_k P^s_m - 
                      P^l_k (V_\mu)^s_m) B_{lsn} \,, \nonumber
\nonumber 
\en 
where $B_{lmn}$ is a tensor with indices $l,m,n$ 
running from 1 to 4 representing the 20--plet of baryons 
(see details in Refs.~\cite{Okubo:1975sc}); $(V_\mu)^l_k$ and 
$P^l_k$ are the matrices representing the 15--plets of vector 
and pseudoscalar fields. The baryon tensor satisfies the conditions 
\eq 
B_{lmn} + B_{mnl} + B_{nlm} = 0, \hspace*{.5cm} 
B_{lmn} = B_{mln} \,. 
\en 
The full list of physical states in terms of SU(4) tensors is 
given in Ref.~\cite{Okubo:1975sc}. Here we only display a few of them: 
\eq 
& &p = B_{112} = - 2 B_{121} = - 2 B_{211}\,, \hspace*{.25cm}
   n = - B_{221} = 2 B_{212} = 2 B_{122}\,, \nonumber\\ 
& &\Sigma_c^{++} = B_{114} = - 2 B_{141} = - 2 B_{411}\,, \hspace*{.25cm}
   \Sigma_c^0    = - B_{224} = 2 B_{242} = 2 B_{422}\,, \\ 
& &\pi^+ = P^2_1\,, \hspace*{.25cm} 
   \pi^- = P^1_2\,, \hspace*{.25cm}  
    D^0  = P^1_4\,, \hspace*{.25cm} 
    D^{\ast +} = V^2_4\,, \hspace*{.25cm}  
    D^{\ast 0} = V^1_4 \,. \nonumber 
\en  
The matching of the SU(3) and SU(4) $VPBB$ Lagrangians at tree level 
gives the following relations between the effective couplings: 
\eq 
g_{\rho\pi\pi} \frac{D}{F_P} = \frac{3}{4} g_1 - \frac{3}{2} g_2 \,, 
\nonumber \\
g_{\rho\pi\pi} \frac{F}{F_P} = \frac{5}{4} g_1 - g_2 \,.  
\en 
Note that the coupling constant $g_3$ is left to be unmatched. 
Below we display the terms of the SU(4) $VPBB$ Lagrangian relevant 
for our calculations: 
\eq 
{\cal L}_{VPBB} &=& \Big( g_2 - \frac{5}{4} g_1 \Big) 
\bar p i \gamma^\mu \gamma^5 p D^{\ast 0}_\mu \bar D^0 
- \Big( g_1 - \frac{5}{4} g_2 \Big) 
\bar p i \gamma^\mu \gamma^5 n D^{\ast +}_\mu \bar D^0 \nonumber\\
&+&\Big( \frac{g_1+g_2}{4} - \frac{3}{2} g_3 \Big) 
\Big( \bar\Sigma_c^{++} i \gamma^\mu\gamma^5 p D^{\ast 0} \pi^+ 
    + \bar\Sigma_c^{0} i \gamma^\mu\gamma^5  n D^{\ast +} \pi^- \Big) \\ 
&-& \frac{3}{2} g_3 
\Big( \bar\Sigma_c^{++} i \gamma^\mu\gamma^5 n D^{\ast +} \pi^+ 
    + \bar\Sigma_c^{0}  i \gamma^\mu\gamma^5 p D^{\ast 0} \pi^- \Big) 
\ + \ {\rm H.c.} \nonumber
\en 
We can estimate the coupling $g_3$ using the following procedure:
the corresponding vertices are generated by 
static one--nucleon exchange 
between the pairs of (nucleon, $\pi$) and ($D^\ast$, $\Sigma_c$) [see Fig.3]. 
We therefore can express the couplings of 
$\bar\Sigma_c^{++} n D^{\ast +} \pi^+$ and 
$\bar\Sigma_c^{0}  p D^{\ast 0} \pi^-$ (which are proportional to the 
coupling $g_3$) in terms of the $\pi NN$ and $D^\ast N \Sigma_c$ couplings as 
\eq 
- \frac{3}{2} g_3 = \frac{g_{D^\ast N \Sigma_c}g_{\pi NN}}{m_N} \sqrt{2} \; .
\en 
Here the coupling $g_{D^\ast N\Sigma_c}$ can be fixed by
the matching of SU(3) and SU(4) $VBB$ Lagrangians. In particular, 
the SU(4) $VBB$ Lagrangian has the form (here and in the following 
we neglect the tensorial part of the $VBB$ interaction containing a
derivative 
acting on the vector field) 
\eq 
{\cal L}_{VBB} = f_1 \bar B^{kmn} \gamma^\mu (V_\mu)_k^l B_{lmn} 
+ f_2 \bar B^{kmn} \gamma^\mu (V_\mu)_k^l B_{lnm}
\en 
where the $f_1$ and $f_2$ are the coupling constants. We do not include 
the term $f_3 \bar B^{kmn} \gamma^\mu B_{kmn} \, {\rm tr}(V_\mu)$ because 
we suppose that the $\phi NN$ coupling vanishes due to 
the Okuba-Zweig-Iizuka (OZI) rule.
In Ref.~\cite{Meissner:1997qt} it was shown  
that an estimate of the ${\phi NN}$ coupling from a dispersive analysis 
results in the value $g_{_{\phi NN}} = - 0.24$. Using the 
definitions of the $\rho NN$ and $\omega NN$ couplings with 
\eq 
{\cal L}_{\rho NN} = \frac{g_{\rho NN}}{2} \bar N \gamma^\mu 
\vec\rho_\mu \vec\tau N \, \nonumber\\
{\cal L}_{\omega NN} = \frac{g_{\omega NN}}{2} \bar N \gamma^\mu 
\omega_\mu N 
\en 
we can express the SU(4) couplings in terms of the $g_{\rho NN}$ and 
$g_{\omega NN}$ coupling constants as: 
\eq 
f_1 = \frac{2}{3\sqrt{2}} \Big( \frac{5}{3} g_{\omega NN} 
- g_{\rho NN} \Big) \,, \nonumber\\
f_2 = \frac{4}{3\sqrt{2}} \Big( \frac{2}{3} g_{\omega NN} 
- g_{\rho NN} \Big) \,. 
\en 
Taking the SU(3) predictions for the $g_{\rho NN}$ and  
$g_{\omega NN}$ couplings of 
\eq 
g_{\rho NN} = 6\,, \hspace*{.25cm} g_{\omega NN} = 3 g_{\rho NN} 
\en 
we get 
\eq 
f_1 = \frac{8}{3\sqrt{2}} g_{\rho NN} \simeq 11.32 \,, \hspace*{.25cm} 
f_2 = \frac{4}{3\sqrt{2}} g_{\rho NN} \simeq  5.66 \,, \hspace*{.25cm} 
g_{D^\ast N \Sigma_c} = \frac{1}{4} \Big( f_1 + f_2 \big) 
= \frac{1}{\sqrt{2}} g_{\rho NN} \simeq 4.24 \,. 
\en 
Finally we get for the SU(4) coupling $g_3$ the expression 
\eq 
g_3 = - \frac{2}{3 F_P} g_{\rho\pi\pi} g_A
\en 
where we use the universality of the $\rho$ meson with 
$g_{\rho\pi\pi} = g_{\rho NN}$. In the evaluation we use 
different values for the leptonic decay constants $F_\pi$, $F_D$, etc. 
associated with $\pi$, $D$, etc. in order to take
into account flavor symmetry breaking corrections.

\newpage 

\begin{figure}
\centering{\
\epsfig{figure=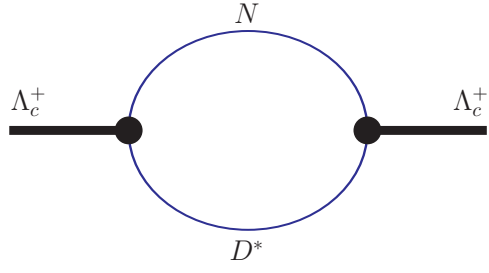,scale=.75}}
\caption{Diagram describing the $\lc$ mass operator.}
\label{fig:str}
\end{figure}

\vspace*{2cm} 

\begin{figure}
\centering{\
\epsfig{figure=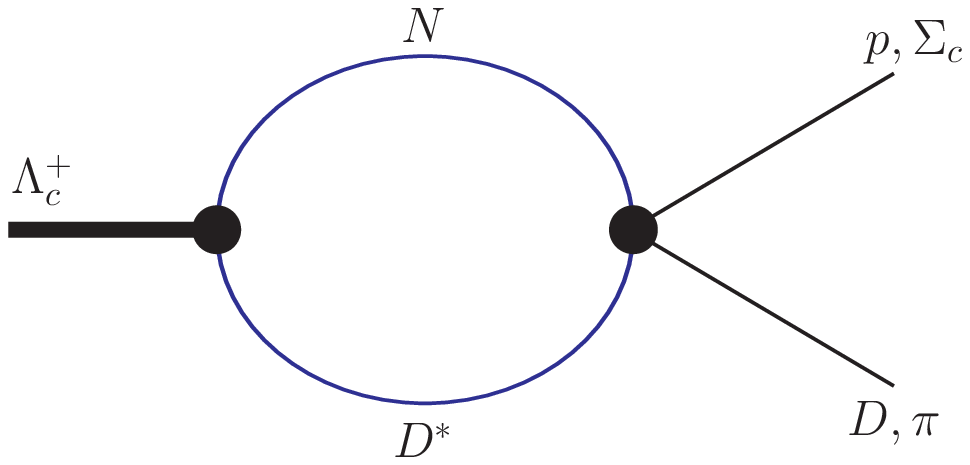,scale=.75}}
\caption{Diagrams contributing to the
decays $\lc \to pD^0, \Sigma_c^{++}\pi^-, \Sigma_c^0 \pi^+$.}
\label{fig:vertex}

\centering{\
\epsfig{figure=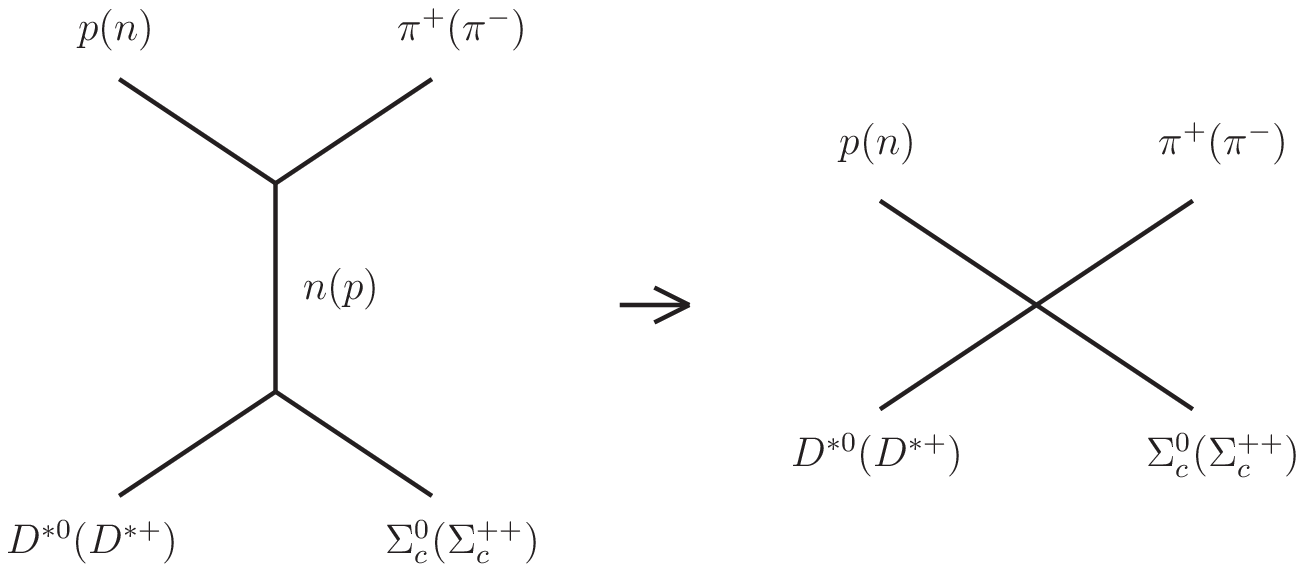,scale=.75}}

\vspace*{3cm}
\caption{Estimate for the coupling $g_3$.}
\label{fig:app}
\end{figure}

\begin{table}
\begin{center}
{\bf Table 1.} 
Partial decay widths of $\lc$ in MeV. 
\vspace*{.25cm}

\def\arraystretch{1}
\begin{tabular}{|c||c|c|c||c|c|c|}  
\hline 
& \multicolumn{3}{c|}{$\frac{1}{2}^+$ modes}
& \multicolumn{3}{c|}{$\frac{1}{2}^-$ modes}\\ 
\hline
$\cos\theta$ & $\Lambda_c^+ \to p D^0$ 
    & $\Lambda_c^+ \to \Sigma_c^{++}\pi^-$  
    & $\Lambda_c^+ \to \Sigma_c^{0} \pi^+$ 
    & $\Lambda_c^+ \to p D^0$ 
    & $\Lambda_c^+ \to \Sigma_c^{++}\pi^-$ 
    & $\Lambda_c^+ \to \Sigma_c^{0} \pi^+$ \\
\hline 
1   & 0.11 & 0.58 & 0.72 & 19.15 & 612.68 & 756.72  \\
0.95& 0.17 & 0.85 & 0.98 & 29.75 & 907.64 & 1040.36  \\
0.9 & 0.20 & 0.96 & 1.08 & 34.40 & 1033.00 & 1153.95 \\
0.8 & 0.23 & 1.11 & 1.20 & 41.09 & 1208.89 & 1305.10 \\
0.7 & 0.25 & 1.20 & 1.27 & 46.17 & 1338.06 & 1407.80 \\
0.6 & 0.27 & 1.27 & 1.30 & 50.24 & 1437.58 & 1478.96 \\
0.5 & 0.28 & 1.31 & 1.32 & 53.47 & 1511.85 & 1522.78 \\
0.4 & 0.29 & 1.32 & 1.30 & 55.83 & 1560.10 & 1538.24 \\
0.3 & 0.29 & 1.32 & 1.30 & 55.83 & 1560.10 & 1538.24 \\
0.2 & 0.29 & 1.30 & 1.26 & 57.15 & 1577.04 & 1519.78 \\
0.1 & 0.26 & 1.14 & 1.03 & 54.20 & 1447.05 & 1309.75 \\
0.05& 0.24 & 1.04 & 0.91 & 50.68 & 1334.05 & 1174.51 \\
0   & 0.18 & 0.74 & 0.60 & 38.15 & 964.41  & 781.52  \\
\hline
\end{tabular}
\end{center}
\end{table} 


\begin{thebibliography}{99} 

\bibitem{Aubert:2006sp} 
  B.~Aubert {\it et al.}  [BABAR Collaboration],
  Phys.\ Rev.\ Lett.\  {\bf 98}, 012001 (2007)
  [arXiv:hep-ex/0603052].

\bibitem{Abe:2006rz}
  R.~Mizuk {\it et al.}  [Belle Collaboration],
  Phys.\ Rev.\ Lett.\  {\bf 98}, 262001 (2007)
  [arXiv:hep-ex/0608043].

\bibitem{He:2006is}
  X.~G.~He, X.~Q.~Li, X.~Liu and X.~Q.~Zeng,
  Eur.\ Phys.\ J.\  C {\bf 51}, 883 (2007)
  [arXiv:hep-ph/0606015].

\bibitem{Capstick:1986bm}
  S.~Capstick and N.~Isgur,
  Phys.\ Rev.\  D {\bf 34}, 2809 (1986); 
  L.~A.~Copley, N.~Isgur and G.~Karl,
  Phys.\ Rev.\  D {\bf 20}, 768 (1979)
  [Erratum-ibid.\  D {\bf 23}, 817 (1981)].

\bibitem{Chen:2007xf}
  C.~Chen, X.~L.~Chen, X.~Liu, W.~Z.~Deng and S.~L.~Zhu,
  Phys.\ Rev.\  D {\bf 75}, 094017 (2007)
  [arXiv:0704.0075 [hep-ph]].

\bibitem{Ebert:2007nw}
  D.~Ebert, R.~N.~Faustov and V.~O.~Galkin,
  Phys.\ Lett.\  B {\bf 659}, 612 (2008)
  [arXiv:0705.2957 [hep-ph]].

\bibitem{Zhong:2007gp}
  X.~H.~Zhong and Q.~Zhao,
  Phys.\ Rev.\  D {\bf 77}, 074008 (2008)
  [arXiv:0711.4645 [hep-ph]].

\bibitem{Cheng:2006dk}
  H.~Y.~Cheng and C.~K.~Chua,
  Phys.\ Rev.\  D {\bf 75}, 014006 (2007)
  [arXiv:hep-ph/0610283].

\bibitem{Gerasyuta:2007un}
  S.~M.~Gerasyuta and E.~E.~Matskevich,
  Int.\ J.\ Mod.\ Phys.\  E {\bf 17}, 585 (2008)
  [arXiv:0709.0397 [hep-ph]].

\bibitem{Roberts:2007ni}
  W.~Roberts and M.~Pervin,
  Int.\ J.\ Mod.\ Phys.\  A {\bf 23}, 2817 (2008)
  [arXiv:0711.2492 [nucl-th]].

\bibitem{GarciaRecio:2008dp}
  C.~Garcia-Recio, V.~K.~Magas, T.~Mizutani, J.~Nieves, 
  A.~Ramos, L.~L.~Salcedo and L.~Tolos,
  Phys.\ Rev.\  D {\bf 79}, 054004 (2009)
  [arXiv:0807.2969 [hep-ph]].  

\bibitem{Valcarce:2008dr}
  A.~Valcarce, H.~Garcilazo and J.~Vijande,
  Eur.\ Phys.\ J.\  A {\bf 37}, 217 (2008)
  [arXiv:0807.2973 [hep-ph]].

\bibitem{Wang:2008vj}
  Q.~W.~Wang and P.~M.~Zhang,
  arXiv:0810.5609 [hep-ph].

\bibitem{Chen:2009tm}
  B.~Chen, D.~X.~Wang and A.~Zhang,
  arXiv:0906.3934 [hep-ph].

\bibitem{Faessler:2007gv}
  A.~Faessler, T.~Gutsche, V.~E.~Lyubovitskij and Y.~L.~Ma,
  Phys.\ Rev.\  D {\bf 76}, 014005 (2007)
  [arXiv:0705.0254 [hep-ph]]; 
  A.~Faessler, T.~Gutsche, S.~Kovalenko and V.~E.~Lyubovitskij,
  Phys.\ Rev.\  D {\bf 76}, 014003 (2007)
  [arXiv:0705.0892 [hep-ph]]; 
  A.~Faessler, T.~Gutsche, V.~E.~Lyubovitskij and Y.~L.~Ma,
  Phys.\ Rev.\  D {\bf 76}, 114008 (2007)
  [arXiv:0709.3946 [hep-ph]]; 
  A.~Faessler, T.~Gutsche, V.~E.~Lyubovitskij and Y.~L.~Ma,
  Phys.\ Rev.\  D {\bf 77}, 114013 (2008)
  [arXiv:0801.2232 [hep-ph]].  
\bibitem{Dong:2008gb}
  Y.~B.~Dong, A.~Faessler, T.~Gutsche and V.~E.~Lyubovitskij,
  Phys.\ Rev.\  D {\bf 77}, 094013 (2008)
  [arXiv:0802.3610 [hep-ph]]; 
  Y.~B.~Dong, A.~Faessler, T.~Gutsche, S.~Kovalenko and V.~E.~Lyubovitskij,
  Phys.\ Rev.\  D {\bf 79}, 094013 (2009)
  [arXiv:0903.5416 [hep-ph]]; 
  Y.~Dong, A.~Faessler, T.~Gutsche and V.~E.~Lyubovitskij,
  arXiv:0909.0380 [hep-ph].
  T.~Branz, T.~Gutsche and V.~E.~Lyubovitskij,
  Phys.\ Rev.\  D {\bf 79}, 014035 (2009)
  [arXiv:0812.0942 [hep-ph]]; 
  T.~Branz, T.~Gutsche and V.~E.~Lyubovitskij,
  Phys.\ Rev.\  D {\bf 80}, 054019 (2009)
  [arXiv:0903.5424 [hep-ph]]. 
\bibitem{Weinberg:1962hj}
  S.~Weinberg,
  Phys.\ Rev.\  {\bf 130}, 776 (1963);
  A.~Salam,
  Nuovo Cim.\  {\bf 25}, 224 (1962);
  K.~Hayashi, M.~Hirayama, T.~Muta, N.~Seto and T.~Shirafuji,
  Fortsch.\ Phys.\ {\bf 15}, 625 (1967).
\bibitem{Efimov:1993ei}
  G.~V.~Efimov and M.~A.~Ivanov,
  {\it The Quark Confinement Model of Hadrons},
  (IOP Publishing, Bristol $\&$ Philadelphia, 1993). 
\bibitem{Anikin:1995cf}
  I.~V.~Anikin, M.~A.~Ivanov, N.~B.~Kulimanova and V.~E.~Lyubovitskij,
  Z.\ Phys.\ C {\bf 65}, 681 (1995);
  M.~A.~Ivanov, M.~P.~Locher and V.~E.~Lyubovitskij,
  Few Body Syst.\  {\bf 21}, 131 (1996); 
  M.~A.~Ivanov, V.~E.~Lyubovitskij, J.~G.~K\"orner and P.~Kroll,
  Phys.\ Rev.\ D {\bf 56}, 348 (1997)
  [arXiv:hep-ph/9612463];
  M.~A.~Ivanov, J.~G.~K\"orner, V.~E.~Lyubovitskij and A.~G.~Rusetsky, 
  Phys.\ Rev.\ D {\bf 60}, 094002 (1999) 
  [arXiv:hep-ph/9904421]; 
  A.~Faessler, T.~Gutsche, M.~A.~Ivanov, V.~E.~Lyubovitskij and P.~Wang, 
  Phys.\ Rev.\  D {\bf 68}, 014011 (2003) 
  [arXiv:hep-ph/0304031];  
  A.~Faessler, T.~Gutsche, M.~A.~Ivanov, J.~G.~Korner,
  V.~E.~Lyubovitskij, D.~Nicmorus and K.~Pumsa-ard,
  Phys.\ Rev.\ D {\bf 73}, 094013 (2006)
  [arXiv:hep-ph/0602193]; 
  A.~Faessler, T.~Gutsche, B.~R.~Holstein, V.~E.~Lyubovitskij,
  D.~Nicmorus and K.~Pumsa-ard,
  Phys.\ Rev.\ D {\bf 74}, 074010 (2006)
  [arXiv:hep-ph/0608015]; 
  A.~Faessler, T.~Gutsche, B.~R.~Holstein, M.~A.~Ivanov, J.~G.~Korner and 
  V.~E.~Lyubovitskij,
  Phys.\ Rev.\  D {\bf 78}, 094005 (2008)
  [arXiv:0809.4159 [hep-ph]]; 
  A.~Faessler, T.~Gutsche, M.~A.~Ivanov, J.~G.~Korner and V.~E.~Lyubovitskij,
  Phys.\ Rev.\  D {\bf 80}, 034025 (2009)
  [arXiv:0907.0563 [hep-ph]].

\bibitem{ChPT}
S.~Weinberg,
Physica A {\bf 96} (1979) 327;
J.~Gasser and H.~Leutwyler,
Annals Phys.\ {\bf 158}, 142 (1984).

\bibitem{Bando:1987br}
  M.~Bando, T.~Kugo and K.~Yamawaki,
  Phys.\ Rept.\  {\bf 164}, 217 (1988); 
  M.~Bando, T.~Kugo, S.~Uehara, K.~Yamawaki and T.~Yanagida,
  Phys.\ Rev.\ Lett.\  {\bf 54}, 1215 (1985).

\bibitem{Okubo:1975sc}
  S.~Okubo,
  Phys.\ Rev.\  D {\bf 11}, 3261 (1975);
  W.~Liu, C.~M.~Ko and Z.~W.~Lin,
  Phys.\ Rev.\  C {\bf 65}, 015203 (2001).

\bibitem{Meissner:1997qt}
U.~G.~Meissner, V.~Mull, J.~Speth and J.~W.~van Orden,
Phys.\ Lett.\ B {\bf 408}, 381 (1997)
[arXiv:hep-ph/9701296].

\end{thebibliography}
\end{document}